\documentclass[aps,prd,article,showpacs]{revtex4}%
\usepackage{amsmath}
\usepackage{graphicx}
\usepackage{epsfig}
\usepackage{amsfonts}
\usepackage{amssymb}%
\providecommand{\U}[1]{\protect\rule{.1in}{.1in}}
\providecommand{\U}[1]{\protect\rule{.1in}{.1in}}
\begin{document}
\title{A stochastic version of the Noether Theorem}
\author{Alfredo Gonz\'{a}lez Lezcano$^{a}$ and Alejandro Cabo Montes de Oca$^{b}$}
\affiliation{$^{a}$ Departamento de F\'{\i}sica, Universidad de Pinar del R\'{\i}o,}
\affiliation{Pinar del R\'{\i}o, Cuba,\\}
\affiliation{$^{b}$ Instituto de Matem\'{a}tica, Cibern\'{e}tica y F\'{\i}sica, }
\affiliation{Calle E, No. 309, e/ 13 y 15, Vedado, La Habana, Cuba. }

\begin{abstract}
\noindent A stochastic version of the Noether Theorem is derived for systems
under the action of external random forces. The concept of moment generating
functional is employed to describe the symmetry of the stochastic forces. The
theorem is applied to two kinds of random covariant forces. One of them
generated in an electrodynamic way and the other defined in the rest frame of
the particle as a function of the proper time. For both of them, it is shown
the conservation of the mean value of a random drift momentum. The validity of
the theorem makes clear that random systems can produce causal stochastic
correlations between two faraway separated systems, that had interacted in the
past. In addition possible connections of the discussion with the Ives
Couder's experimental results are remarked.

\end{abstract}

\maketitle

%\date{}

\section{Introduction}

\label{intro} The Noether theorem is an important result in theoretical
physics and can be roughly stated as: If the Lagrangian of a physical system
has a certain symmetry, then there is a conservation law for a quantity
$\mathcal{J}$ called as the Noether's current
\cite{Bogo,Goldstein,Sokolov,Landau}. Depending of the symmetry of the
Lagrangian, the Noether current becomes angular momentum, linear momentum and
energy, for spatial isotropy, spatial homogeneity and time invariance,
respectively. Consider the dynamic system described by a Lagrangian
$\mathcal{L}$ with certain symmetries. Then, according to the  theorem, the
corresponding  currents will be conserved. In the presence of a generic
external forces, this conservation law is not valid in general. If these external
forces has some random behavior, then the question arise of whether it is
possible to keep standing a conservation law at least for means values of the
Noether currents. There are some situations where the influence of the
stochastic field on the system is in certain sense symmetric. That is, for
instance, in case that the spatial homogeneity, if somehow the field has the
same influence in different points of the space. Stochastic forces are
functions of the form $Q(q,\overset{\cdot}{q},\theta)$
\cite{Vankampen,Caceres}, being $q$, $\overset{\cdot}{q}$ and $\theta$ a
compact notation for a set of generalized coordinates, velocities and random
variables, respectively. In reference  \cite{Alejandro} it was suggested the
possibility of an extension of the Noether Theorem to a system of particles
subject to the action of an external stochastic field. Simulations of one
particle trajectories (and binary collisions for interacting particles)
predicted the conservation of mean values for linear momentum. This conclusion
was a considered to be a consequence of the relativistic invariance of the
external force, which was defined in that work as a function of the proper time. The
idea of the conservation of mean values of linear momentum for particles
submitted to external forces was also suggested by a series of experiments
recently carried out by Couder \textit{etal} \ \cite{Couder1,Couder2,Couder3}.
In these works a situation consisting of a liquid droplet bouncing over a
vibrating liquid surface was experimentally investigated. The measurements
showed that the interaction of the droplet with the surface wave it creates,
causes (under some specific experimental conditions) a drift movement with
conserved mean velocity. The experiments also exhibited interesting properties
strongly resembling the quantum behavior. The possibility of an extension of
the Noether Theorem for systems sharing the main characteristic of being
affected by stochastic forces is closely related with the symmetry properties
of those forces. To speak about symmetry of a stochastic
field could bring about some confusion, taking into account that the used
transformations do not affect the random variables. That is: to impose
symmetry conditions to the these forces can't be done directly on the function
$Q(q,\overset{\cdot}{q},\theta)$ because the stochastic character of the field
avoids the possibility to predict the value of the variation for different
settings of $\theta$. In order to be able to make a general extension of the
Noether Theorem for this kind of systems, it will be employed a functional
that, in analogy with the action of dynamic systems
\cite{Bogo,Goldstein,Sokolov,Landau}, can fully describe the stochastic process
defining the forces and then impose symmetry conditions to this functional.  It should remarked
that a slightly different formulation of a stochastic Noether theorem was
advanced in reference \cite{baez}. This work, basically discusses the
cases of special Markov's random processes which seems to be much restricted
\ than the here considered ones. The only  constraint of the discussion here is 
that the physical system has equations of motions coming from a Lagrangian. 
Thus, we consider our  discussion   as being  complementary  to the one done in \cite{baez}.

The analysis in general supports the possibility that two random classical
systems which interact during a finite time lapse, and afterwards flight far
apart, can retain correlations between their physical properties, which can
describe apparent non causal links between physical quantities measured in the
two well separated classical systems, after they stop to interact. This
property seems to leave space for the explanation of EPR effects (\cite{EPR})
in hidden variable theories, in a way  satisfying the Bell restrictions about the
existence of such models \cite{bell}.

\ It also suggests the interest of coupling the "self-field" to stochastically
driven particles as the ones discussed here. That is, to consider the existence  at
the particle place a point sources of the Klein-Gordon field, by example.
 In these cases, let us assume that a momentum conservation can be derived. Then, if a system of reference
exists in which the momentum vanishes and the coupled particle-wave-modes show a
spacially localized structure, it will imply that in this frame,  the random
distribution might describe a spatially localized stationary movements of the
particle coupled with the stochastically populated self-field  modes. But,
assuming that the system is relativistic invariant, in a different Lorentz frame, they  will
exist analogous, but uniformly moving localized structures. \ Therefore, the
situation is quite resembling the one in the mentioned Couder's experiments
\cite{Couder1,Couder2,Couder3}. If these expected particle wave composite
structures could show interference effects when approaching two slits in a
wall, a connection of the Couder's experimental results with microscopic
hidden variable structures could be found. We expect to consider these
possibilities in extensions of this work.

In Section 2 the stochastic form of the Noether theorem is formulated and
proved for a general class of stochastic forces. Next, Sections 3 and 4,
considers its application to the above mentioned two physical systems, in
proving that they allow for motions conserving a mean drift velocity of the
particles. In the Summary the results are reviewed and few remarks on the
possible extensions of the work are advanced.

\section{Stochastic Noether theorem}

Let us consider a the dynamic system described by a Lagrangian $L$
depending of $N$ generalized coordinates $q_{i},$ \ $i=1,...,N$ and their
respective velocities $\overset{\cdot}{q},$ in which the point indicates the
time derivative. For considering the presence of an external force acting on
the system, we will use the recourse of including in the Lagrangian a term
from which the force is derived. However, to find the Lagrangian terms
determining a force is not always easy or possible. The search of those action
terms is a difficult problem and it is not satisfactorily solved for a generic
force. However, immediate and exacts solutions are known for the
electromagnetic forces and many relevant physical systems. We will restrict
our discussion to those cases.

Nevertheless, we will start the discussion without initially requiring that
the forces are derived from a Lagrangian. Then, in general, the Euler-Lagrange
equations of motion can be written in the form:%
\begin{equation}
\frac{\partial L}{\partial q_{i}}-\frac{d}{d\tau}\left(  \frac{\partial
L}{\partial\dot{q_{i}}}\right)  =-Q_{i}\left(  q_i,\dot{q_{i}}\right)  ,
\label{Elag}%
\end{equation}
where $Q_{i}\left(  q_i,\dot{q_{i}}\right)  $ are the generalized forces that
act on the system and depend on both the coordinates and velocities. These
explicit dependencies will be in general omitted for simplifying the notation.
If the unperturbed by the force system has a certain symmetry, then the
variation of the Lagrangian with respect to the change in the variables
defined by the symmetry should vanish
\begin{equation}
\delta L=\sum_{i}(\frac{\partial L}{\partial q_{i}}\delta q_{i}+\frac{\partial
L}{\partial\dot{q_{i}}}\delta\dot{q_{i}})=0. \label{Variation}%
\end{equation}

In (\ref{Variation}) $\delta q_{i}$ and $\delta\dot{q_{i}}$ are just the
changes in the generalized coordinates, the variation of which makes the
Lagrangian to remain invariant. Now, lets us consider the standard reasonings
for the demonstration of the Noether theorem in usual situations \cite{Bogo},
but in this case in connection with the inhomogeneous Euler-Lagrange equations
(\ref{Elag}). Thus, making use of the derivative of a product in
(\ref{Variation}) and also considering that the coordinate variations are done
at fixed instant of time, that is:
\[
\delta\dot{q_{i}}=\frac{\partial}{\partial t}\delta q_{i},
\]
leads to:
\begin{equation}
\sum_{i}\left(  \left[  \frac{\partial L}{\partial q_{i}}-\frac{d}{d\tau
}\left(  \frac{\partial L}{\partial\dot{q_{i}}}\right)  \right]  \delta
q_{i}+\frac{d}{d\tau}\left(  \frac{\partial L}{\partial\dot{q_{i}}}\delta
q_{i}\right)  \right)  =0.
\end{equation}
Then, from (\ref{Elag}) results:
\begin{equation}
\frac{d}{d\tau}\sum_{i}\left(  \frac{\partial L}{\partial\dot{q_{i}}}\delta
q_{i}\right)  =\sum_{i}Q_{i}\delta q_{i}. \label{Conservation}%
\end{equation}
If the symmetry group $G$ associated to the unperturbed Lagrangian (the one
omitting the random forces) has an infinitesimal transformation generated by
$r_{max}$ parameters $r=1,2,...,r_{\max}$, the variations can be expressed as
\[
\delta q_{i}=\sum_{r}\varepsilon^{r}\Lambda_{i}^{r}\left(  q\right)  ,\text{ }%
\]
and their substitution in (\ref{Conservation}) leads to:
\begin{equation}
\frac{d}{d\tau}\sum_{i}\left(  \frac{\partial L}{\partial\dot{q}_{i}%
}\varepsilon^{r}\Lambda_{i}^{r}\left(  q\right)  \right)  =\sum_{i,r}%
Q_{i}\text{ }\varepsilon^{r}\Lambda_{i}^{r}\left(  q\right)  .
\end{equation}

Assuming the independence of the parameters $\varepsilon^{r}$, it is obtained
that:
\begin{align}
\frac{d\mathcal{J}\text{ }^{r}}{d\tau}  &  =\sum_{i}Q_{i}\text{ }\Lambda
_{i}^{r}\left(  q\right)  ,\text{ \ \ \ }r=1,...,r_{max}, \label{conservation}%
\\
\mathcal{J}\text{ }^{r}  &  =\sum_{i}\frac{\partial L}{\partial\dot{q}_{i}%
}\Lambda_{i}^{r}\left(  q\right)  ,
\end{align}
where $\mathcal{J}^{\text{ }r}=\sum_{i}\frac{\partial L}{\partial\dot{q}_{i}%
}\Lambda_{i}^{r}\left(  q\right)  $ for the various values of the index $r,$
are the Noether currents which will be conserved if the generalized forces
vanish. Further, we will assume that the stochastic forces are derivable from
a Lagrangian $L^{ram}$ as
\begin{align*}
Q_{i}(q_i,\dot{q})\text{ }  &  \text{=}\frac{\delta S^{ram}[q_i,\dot{q_{i}}%
]}{\delta q_{i}(t)}=-\left(  \frac{\partial L^{ram}(q_i,\dot{q_{i}})}{\partial
q_{i}(t)}-\frac{\partial}{\partial t}(\frac{\partial L^{ram}(q_i,\dot{q_{i}}%
)}{\partial\dot{q_{i}}(t)})\right)  ,\\
S^{ram}[q_i,\dot{q_{i}}]  &  =\int dt\text{ \ }L^{ram}(q_i,\dot{q_{i}}).
\end{align*}

\subsection{The random \ forces}

Since we will discuss the situation in which the forces are random, let us now
define those forces. The aleatory behavior will be introduced by the
dependence of the forces not only of the generalized coordinates
$q=\{q_{1},q_{2},...,q_{i},...,q_{N}\}$ and their velocities, but also on a
set of random variables $\theta\in\Theta$, being $\Theta$ a generic sample
space. That is $Q_{i}=Q_{i}(q,\dot{q},\theta).$ Then, assuming well defined
initial conditions for the set of generalized coordinates $q(t_{0})=q_{0}$ at
an initial fixed instant of time $t_{0},$ we can define an ensemble of a large
number of coordinate trajectories $q(t,\theta)$ all satisfying the initial
conditions but following different courses as the time increases in dependence
of the random value of the parameter $\theta.$ Therefore, we will define a
mean value of an arbitrary function $R(q_{1}(t,\theta),,...,q_{N}(t,\theta)|$
$\dot{q}_{1}(t,\theta),...,\dot{q}_{N}(t,\theta))$ of a set of coordinates
and their velocities taken at a fixed instant $t$, in the form
\begin{equation}
\langle\text{ }R(q_{1}(t,\theta),...,q_{N}(t,\theta)|\text{ }\dot{q}%
_{1}(t,\theta),...,\dot{q}_{N}(t,\theta))\text{ }\rangle_{\theta}=\int
d\theta\text{ }\rho(\theta)\text{ }R(q_{1}(t,\theta),...,q_{N}(t,\theta
)|\text{ }\dot{q}_{1}(t,\theta),...,\dot{q}_{N}(t,\theta)))\text{,}\label{R}%
\end{equation}
where the integration runs over the space of parameters $\theta$ and
$\rho(\theta)$ is the probability density in this space. It is also possible
to define, for an arbitrary time $t$, the density of the ending values of the
coordinates and velocities in the phase space of coordinates and velocities.
For this purpose, consider the large number of $\theta$ values $N_{\theta}$
and the number of ending points $N_{box}(t)$ of the coordinates and velocities
$(q_{i},\overset{\cdot}{q}_{i})$ which at the time $t$, are inside a box of
the phase space of size $\prod_{i}\Delta q_{i}\Delta\overset{\cdot}{q}_{i}%
.$ Then, the density of points in phase space can be defined as
\begin{equation}
\rho(q_{1}(t,\theta),..,q_{N}(t,\theta)\text{ ; }\dot{q}_{1}(t,\theta
),...,\dot{q}_{N}(t,\theta))=\frac{N_{box}(t)}{N_{\theta}}.
\end{equation}
This definition allows to express the mean value (\ref{R}) in an alternative
form as an integral of the quantity $R$ multiplied by \ the density in phase
space $\ \rho$ as
\begin{align}
\langle\text{ }R\text{ }\rangle_{_{\theta}} &  =\int d\theta\rho(\theta)\text{
}R(q_{1}(t,\theta),..,q_{N}(t,\theta)\text{ ; }\dot{q}_{1}(t,\theta
),...,\dot{q}_{N}(t,\theta))\\
&  =\prod_{i}(\int(dq_{i}d\overset{\cdot}{q}_{i})\rho(q_{1}(t,\theta
),...,q_{N}(t,\theta)\text{ ; }\dot{q}_{1}(t,\theta),...,\dot{q}_{N}%
(t,\theta))R(q_{1}(t,\theta),...,q_{N}(t,\theta)\text{ ; }\dot{q}_{1}%
(t,\theta),...,\dot{q}_{N}(t,\theta)).
\end{align}

Now, if we assume that the generalized forces $Q_{i}$ are of the random kind
defined above $Q_{i}=Q_{i}(q,\dot{q},\theta)$ and take the mean value in both
sides of the equation (\ref{conservation}), it directly follows the relation
\begin{equation}
\frac{d\langle\mathcal{J}\text{ }^{r}\rangle_{_{\theta}}}{dt}=\langle\sum
_{i}Q_{i}(q(t,\theta),\dot{q}(t,\theta),\theta)\text{ }\Lambda_{i}^{r}\left(
q(t,\theta)\right)  \rangle_{_{\theta}}. \label{cond}%
\end{equation}
Therefore, the mean value of the quantity $\sum_{i}Q_{i}(q)$ $\Lambda_{i}%
^{r}\left(  q\right)  $ is equal to the time variation of the Noether current
of the unperturbed Lagrangian. In what follows we will consider the proof of
the Noether theorem and the finding of a formula for their conserved currents
by searching for a connection of the condition (\ref{cond}) with the symmetry
properties of the random force distribution.

For the sake of definiteness let us precisely formulate the random Noether
Theorem as follows:

\textit{If the stochastic forces derivable from a Lagrangian term share a
certain symmetry with the Lagrangian of the system on which they act, then a
conservation law of the stochastic mean value of a Noether current as a
function of time is valid. }

First of all, we must define what will be understood by a symmetry of the
random forces. For this purpose we will make use of the property that a
stochastic process is completely characterized by its moment (or
characteristic) generating functional of the auxiliary variables
$k=\{k_{1}(t),k_{2}(t),...,k_{i}(t),...,k_{N}(t)\}$\cite{Vankampen,Caceres}:
\begin{equation}
\Gamma\left[  k\right]  =\langle exp\left(  i\intop_{-\infty}^{\infty
}k(t)\text{ }L^{ram}\left(  q(t,\theta),\dot{q}(t,\theta),\theta\right)
dt\right)  \rangle_{_{\theta}}, \label{G}%
\end{equation}
where $L^{ram}$ is the previously defined action term generating the random
forces. Then, we will consider  that the symmetry of this functional over the
changes of coordinate variables defined by the symmetry group is equivalent to
the invariance of the corresponding stochastic force \cite{Vankampen,Caceres}.
Thus, if the symmetry group of $G$ is parametric with infinitesimal
transformation $\delta q_{i}=\sum_{r}\varepsilon^{r}\Lambda_{i}^{r}\left(
q\right)  , $ $\ r=1,...,r_{\max}$, then the increment of $\Gamma$ should
exactly vanish when the symmetry coordinate change is done in (\ref{G}) for
arbitrary values of the auxiliary parameters $k$ and the group infinitesimal
constants $\varepsilon^{r}$. Therefore
\begin{align}
\delta\Gamma\left[  k\right]   &  =\langle i\intop_{-\infty}^{\infty
}k(t^{\prime})\text{ }\delta L^{ram}\left(  q(t^{\prime},\theta),\dot
{q}(t^{\prime},\theta),\theta\right)  dt^{\prime}\times\nonumber\\
&  exp\left(  i\intop_{-\infty}^{\infty}k(t^{\prime})\text{ }L^{ram}\left(
q(t^{\prime},\theta),\dot{q}(t^{\prime},\theta),\theta\right)  dt^{\prime
}\right)  \rangle_{_{\theta}}\nonumber\\
&  ={\LARGE \langle}i\intop_{-\infty}^{\infty}k(t^{\prime})\sum_{i}%
(\frac{\partial L^{ram}}{\partial q_{i}(t^{\prime},\theta)}\delta q_{i}%
+\frac{\partial L^{ram}}{\partial\dot{q}_{i}(t^{\prime},\theta)}\frac
{d}{dt^{\prime}}\delta q_{i})dt^{\prime}\times\nonumber\\
&  exp\left(  i\intop_{-\infty}^{\infty}k(t^{\prime})\text{ }L^{ram}\left(
q(t^{\prime},\theta),\dot{q}(t^{\prime},\theta),\theta\right)  dt^{\prime
}\right)  {\LARGE \rangle_{_{\theta}}}\nonumber\\
&  ={\LARGE \langle}i\intop_{-\infty}^{\infty}\sum_{i,r}{\Huge (}k(t^{\prime
}){\large (}\frac{\partial L^{ram}}{\partial q_{i}(t^{\prime},\theta)}%
-\frac{d}{dt}(\frac{\partial L^{ram}}{\partial\dot{q}_{i}(t^{\prime},\theta
)}){\large )}\varepsilon^{r}\Lambda_{i}^{r}\left(  q(t^{\prime},\theta\right)
)+\nonumber\\
&  \frac{d}{dt^{\prime}}{\LARGE (}k(t^{\prime})\sum_{i,r}\frac{\partial\text{
}L^{ram}}{\partial\dot{q}_{i}(t^{\prime},\theta)}\varepsilon^{r}\Lambda
_{i}^{r}\left(  q(t^{\prime},\theta\right)  {\LARGE )}{\Huge )}dt^{\prime
}\times\nonumber\\
&  exp\left(  i\intop_{-\infty}^{\infty}k(t^{\prime})\text{ }L^{ram}\left(
q(t^{\prime},\theta),\dot{q}(t^{\prime},\theta),\theta\right)  dt^{\prime
}\right)  {\LARGE \rangle}\nonumber\\
&  =0. \label{functionalder}%
\end{align}

Then, after assuming that the text functions $k(t)$ tend to vanish when $t$
tends to $\pm\infty$, the following relations follow as conditions for the
stochastic Lagrangian to be symmetric%
\begin{align}
&  {\LARGE \langle}\sum_{i,r}(\frac{\partial L^{ram}}{\partial q_{i}%
(t,\theta)}-\frac{d}{dt}\frac{\partial L^{ram}}{\partial\dot{q}_{i}(t,\theta
)})\Lambda_{i}^{r}\left(  q(t,\theta\right)  ){\LARGE \rangle}_{\theta
}\nonumber\\
&  ={\LARGE \langle}\sum_{i,r}Q_{i}(q(t,\theta),\dot{q}(t,\theta
),\theta)\Lambda_{i}^{r}\left(  q(t,\theta)\right)  {\LARGE \rangle}_{\theta
}=0,\label{conditions}\\
r  &  =1,...,r_{\max}.
\end{align}

But, substituting (\ref{conditions}) in (\ref{cond})%
\begin{align}
\frac{d\langle\mathcal{J}\text{ }^{r}\rangle_{_{\theta}}}{dt}  &  =\langle
\sum_{i}Q_{i}(q(t,\theta),\dot{q}(t,\theta),\theta)\text{ }\Lambda_{i}%
^{r}\left(  q(t,\theta)\right)  \rangle_{_{\theta}}\\
&  =0.
\end{align}

This expression indicates that the Noether currents $\langle\mathcal{J}$
$^{r}\rangle_{_{\theta}}$ are conserved in time
\begin{align}
\frac{d}{dt}\ {\Large \langle}\mathcal{J}^{r}{\LARGE \rangle_{_{\theta}}}  &
=0,\\
\mathcal{J}\text{ }^{r}\mathcal{(}q(t,\theta),\dot{q}(t,\theta))  &  =\sum
_{i}\frac{\partial L\mathcal{(}q(t,\theta),\dot{q}(t,\theta)\mathcal{)}%
}{\partial\dot{q}_{i}(t,\theta)}\Lambda_{i}^{r}\left(  q(t,\theta)\right)  .
\label{lagrancurr}%
\end{align}

This result finishes the proof of the stochastic Noether theorem and the
finding of the formulae for the conserved currents.

\section{Random electrodynamic forces}

Let us consider in this section the application of the former discussion to
stochastic electrodynamic forces. This kind of forces had been considered in
the literature within the theory known as Stochastic Quantum Electrodynamics
(SQED) in which the vacuum is assumed to create a force on a charged particle
which is generated by a specially defined stochastic "vacuum" electromagnetic
field \cite{marshall,Boyer}. To start, below we will give the definition of
the relativistic stochastic process under consideration. The random character
of the forces will be implemented as in the past section, by defining a large
ensemble of particle trajectories space $\{x(t,\theta)\}$ were $x(t,\theta)$
is the position four vector of the particle as a function of time and the
parameter $\theta.$ Changing $\theta$ pertaining to the set $\Theta$ will
produce the ensemble random trajectories $C(\theta).$ As it was mentioned, in
this $\sec$tion the symbol $x$ will indicate a compact form for a four-vector
$x^{\mu}=(x^{_{0}},\overrightarrow{x})$, the time $t=x^{_{0}}$ will be
measured in $cm$ and will be equal to the product of the light velocity in
$cm/s$ and the time measured in seconds. Let us write now a Lagrangian from
which the random equations of motions can derived in the form%
\begin{align}
L_{T}  &  =-m\sqrt{1-(\overrightarrow{v}(t,\theta))^{2}}-q(\phi
(\overrightarrow{x}(t,\theta),t)-\overrightarrow{v}(t,\theta)\cdot
\overrightarrow{A}(\overrightarrow{x}(t,\theta),t)=L+L_{ram},\\
L  &  =-m\sqrt{1-(\overrightarrow{v}(t,\theta))^{2}},\text{ \ \ \ \ \ \ \ }%
L_{ram}=-q(\phi(\overrightarrow{x}(t,\theta),t)-\overrightarrow{v}%
(t,\theta)\cdot\overrightarrow{A}(\overrightarrow{x}(t,\theta),t),\\
\overrightarrow{v}(t,\theta)  &  =\frac{d}{dt}\overrightarrow{x}%
(t,\theta),\text{ \ \ \ \ \ }x(t,\theta)=(t,\overrightarrow{x}(t,\theta
)),\text{ \ \ \ \ \ \ \ \ }t=x_{0}.
\end{align}

Each of the trajectories in the ensemble will be defined by a solution of the
equation of motion given by the above Lagrangian \cite{Boyer}
\begin{align}
\frac{d}{dt}p^{\mu}(t,\theta)  &  =\frac{q}{m}F_{\nu}^{\mu}(x(t,\theta
))\frac{d}{dt}x^{\nu}(t,\theta),\\
p^{\nu}(t,\theta)  &  =m\text{ }u^{\mu}(t,\theta)=m\frac{d}{d\tau}x^{\nu
}(t,\theta)\nonumber\\
&  =\frac{m}{\sqrt{1-(\overrightarrow{v}(t,\theta))^{2}}}\frac{d}{dt}x^{\nu
}(t,\theta)\nonumber\\
&  =\frac{m}{\sqrt{1-(\overrightarrow{v}(t,\theta))^{2}}}\left(
\begin{tabular}
[c]{l}%
$\ \ 1$\\
$\overrightarrow{v}(t,\theta)$%
\end{tabular}
\ \right)  \text{ }.
\end{align}

It should be noted that contracting these equations with $p^{\mu}(t)$ and
using the anti-symmetry of the intensity tensor it follows
\[
p_{\mu}(t)\frac{d}{dt}p^{\mu}(t)=\frac{d}{dt}p^{2}(t)=0,
\]
which is a dynamic constraint. This implies that only three equations from the
four ones are independent. That is, the temporal  equation can be derived form the other
three spatial ones%
\begin{align}
\frac{d}{dt}p^{i}(t,\theta)  &  =\frac{q}{m}(F_{j}^{i}(x(t,\theta),t)\frac
{d}{dt}x^{j}(t,\theta)+F_{0}^{i}(x(t,\theta),t),\\
i  &  =1,2,3.\nonumber
\end{align}
The vector potential $A_{\beta}(x)$ and its field intensity $F_{\gamma\beta}$
are given by%
\begin{align}
F_{\gamma\beta}(x)  &  =\partial_{\gamma}A_{\beta}(x)-\partial_{\beta
}A_{\gamma}(x),\\
A_{\beta}(x)  &  =(0,\overrightarrow{A}(x)),\\
\overrightarrow{A}(x)  &  =\sum_{\lambda=1}^{2}\int d\overrightarrow{k}%
\frac{1}{w_{k}}\overrightarrow{\epsilon}(\overrightarrow{k},\lambda
){h(\overrightarrow{k},\lambda)\times}\nonumber\\
&  \sin(\overrightarrow{k}\cdot\overrightarrow{x}-w_{k}\text{ }x^{0}%
+\theta(\overrightarrow{k},\lambda)),\\
w_{k}  &  =|\overrightarrow{k}|,
\end{align}
where the electromagnetic  field was chosen in the Coulomb gauge 
\begin{equation}
\overrightarrow{\nabla
}\cdot\overrightarrow{A}(x)=0,  \,\, \, \,\phi(x)=0,
\end{equation}
and the $\overrightarrow{\epsilon}(\overrightarrow{k}%
,\lambda)$ are two unit polarization vectors associated to the wave vector
$\overrightarrow{k}$ and satisfying%
\begin{equation}
\overrightarrow{\epsilon}(\overrightarrow{k},\lambda)\cdot
\overrightarrow{\epsilon}(\overrightarrow{k},\lambda^{\prime})=\delta
_{\lambda\lambda^{\prime}},\text{ \ \ \ }\overrightarrow{k}\cdot
\overrightarrow{\epsilon}(\overrightarrow{k},\lambda)=0.
\end{equation}
The number $h$ is chosen as the value determining that each wave mode has a
exactly one half of photon quantum energy $w_{k}:$ $\pi^{2}h^{2}=\frac{1}%
{2}w_{\overrightarrow{k}}.$ The metric tensor in this section is assumed as
diagonal and having $-g_{00}=1=g_{11}=g_{22}=g_{33}.$

Further, the parameters $\theta\in\Theta$ are defined as the set of
random phase functions $\theta=\{\theta(\overrightarrow{k},\lambda)\}$ (one
for each pair of momentum and polarization $(\overrightarrow{k},\lambda)$)
which are uniformly distributed in the interval $(-\pi,\pi)$ \cite{Boyer}.
Then, the distribution function in the space of parameters and the volume
deferential in this space will be defined by the following products
\begin{align}
\rho(\theta)  &  =\prod_{(\overrightarrow{k},\lambda)}\frac{\Theta(\pi
-\theta(\overrightarrow{k},\lambda))\Theta(\theta(\overrightarrow{k}%
,\lambda)-\pi)}{2\pi},\\
d\text{ }\theta &  =\prod_{(\overrightarrow{k},\lambda)}d\text{ }%
\theta(\overrightarrow{k},\lambda).
\end{align}

For the mean value of any function of the coordinates and velocities taken at
a given time instant we write
\begin{equation}
\langle\text{ }R(\overrightarrow{x}(t,\theta)|\text{ }\overrightarrow{v}%
(t,\theta)\text{ }\rangle_{_{\theta}}=\int d\theta\text{ }\rho(\theta)\text{
}R(\overrightarrow{x}(t,\theta)|\text{ }\overrightarrow{v}(t,\theta)\text{
})\text{.}%
\end{equation}

\subsection{ Invariance under space translation}

Let us no consider the invariance transformation
\[
x^{i}\rightarrow x^{i}+\delta x^{i},\ \ \ \ \delta x^{i}=\epsilon^{i},
\]
where the $\epsilon^{i}$ are constant infinitesimal increments in the
coordinates. It is clear that the instantaneous Lagrangian is not invariant
under this transformations. However, the stochastic system, if its random
properties do not depend on the spatial point in which are measured, should
show a special form of this invariance. Let us argue that is the case, as
indicated by the Noether theorem presented in the previous section. This form
of invariance can expressed in the notation of the definitions of the previous
sections with $i=r$ for $r=1,2,3$ $,$ that is $\ q_{i}=x^{i},$ $i=1,2,3.$
Thus, for the function $\Lambda$ generating the transformation it follows
\begin{equation}
\Lambda_{i}^{r}(q)=\delta_{i}^{r}\text{ \ \ },\text{ }\ \ \ \ \ \ \ \ i=1,2,3.
\end{equation}

The unperturbed $L$ and the random $L_{ram}$ Lagrangians can be written as
\begin{align}
L(\overrightarrow{x}(t,\theta),\overrightarrow{v}(t,\theta))  &
=-m\sqrt{1-(\overrightarrow{v}(t,\theta))^{2}},\\
\text{\ }L_{ram}  &  =q\frac{d}{dt}\overrightarrow{x}(t,\theta)\cdot
\overrightarrow{A}(x(t,\theta).
\end{align}

Thus, the possibly conserved currents associated to the Noether theorem take
the forms
\begin{align}
\mathcal{J}\text{ }^{r}\mathcal{(}\overrightarrow{x}(t,\theta
),\overrightarrow{v}(t,\theta))  &  =\frac{\partial L}{\partial\frac{d}%
{dt}x^{i}(t,\theta)}\nonumber\\
&  =\frac{m\text{ }v^{i}(t,\theta)}{\sqrt{1-(\overrightarrow{v}(t,\theta
))^{2}}}, \label{noether}%
\end{align}

Therefore, the satisfaction of the property derived in the previous section
for the stochastic forces to be translation invariant will effectively define
wether or not this charges will be conserved after taken their random mean
values. These conditions take the specific forms
\begin{align}
&  {\LARGE \langle}\sum_{i}\left(  \frac{\partial L^{ram}}{\partial
x^{i}(t,\theta)}-\frac{d}{dt}\frac{\partial L^{ram}}{(\frac{d}{dt}%
x^{i}(t,\theta))}\right)  {\LARGE \rangle}_{\theta}\delta^{ir}\nonumber\\
&  =-{\LARGE \langle}\sum_{i}Q_{i}(\overrightarrow{x}(t,\theta
),\overrightarrow{v}(t,\theta)){\LARGE \rangle_{_{\theta}}}\delta
^{ir}=0,\text{ \ \ \ \ \ }r=1,2,3,
\end{align}
in which the generalized forces have the formulae
\begin{align*}
Q^{i}(\overrightarrow{x}(t,\theta),\overrightarrow{v}(t,\theta))  &  =-\left(
\frac{\partial L^{ram}(\overrightarrow{x}(t,\theta),\overrightarrow{v}%
(t,\theta))}{\partial x_{i}(t,\theta)}-\frac{\partial}{\partial t}%
(\frac{\partial L^{ram}(\overrightarrow{x}(t,\theta),\overrightarrow{v}%
(t,\theta))}{\partial\overset{\cdot}{x_{i}}(t,\theta)})\right) \\
&  =\frac{q}{m}\left(  F_{j}^{i}(\overrightarrow{x}(t,\theta
))\overrightarrow{v}(t,\theta)+F_{0}^{i}(\overrightarrow{x}(t,\theta)\right),
\\
\text{\ }L_{ram}  &  =q\overrightarrow{v}(t,\theta)\cdot\overrightarrow{A}%
(\overrightarrow{x}(t,\theta),t).
\end{align*}

These conditions can be explicitly written as%
\begin{equation}
{\LARGE \langle}Q^{i}(x(t,\theta),v^{j}(t,\theta)){\LARGE \rangle}_{\theta
}={\LARGE \langle}\frac{q}{m}\left(  F_{j}^{i}(x(t,\theta))v^{j}%
(t,\theta)+F_{0}^{i}(x(t,\theta)\right)  {\LARGE \rangle}_{\theta}=0,\text{
}\ \text{\ \ \ }i=1,2,3. \label{condit1}%
\end{equation}
in which, the field intensities are expressed as linear superpositions of
space and time derivatives of the potential $A.$ Thus, when the mode expansion
of the potential is substituted, the result is a superposition of harmonic
functions one for each mode. But, the time and spatial derivatives of those
harmonic functions do not introduce phase functions outside their arguments.
In this situation, let us assume that all but one of the phase functions are
fixed and the integral over the single varying one is considered in
(\ref{condit1}). Then, since the number of modes is large, tending to
infinity, the variation of a single phase, let say $\theta(\overrightarrow{k}%
_{o},\lambda_{o})$ should determine an infinitesimal variation of the
trajectories and velocities $\overrightarrow{x}(t,\theta),\overrightarrow{v}%
(t,\theta).$ Thus, the integral over the phase $\theta(\overrightarrow{k}%
_{o},\lambda)$ in the whole interval $(-\pi,\pi)$ will vanish because \ the
integral \ in a period of any harmonic function is zero. The above argue can
be alternatively illustrated by representing the measure of the random mean
value as the product of the large number of phases which are kept constant and
the integral over the single phase $\theta(\overrightarrow{k}_{o},\lambda
_{o})$
\begin{gather}
\left(  \prod_{(\overrightarrow{k^{\prime}},\lambda^{\prime})}d\text{ }%
\theta(\overrightarrow{k^{\prime}},\lambda^{\prime})\text{ }\Theta(\pi
-\theta(\overrightarrow{k^{\prime}},\lambda^{\prime}))\Theta(\theta
(\overrightarrow{k^{\prime}},\lambda^{\prime})-\pi)\right)  \times\nonumber\\
\int_{-\pi}^{\pi}d\theta(\overrightarrow{k}_{o},\lambda_{o})\text{ }%
\sin(\overrightarrow{k}\cdot\overrightarrow{x}(t,\theta)-w_{k}\text{ }%
x^{0}+\theta(\overrightarrow{k_{o}},\lambda_{o})).
\end{gather}
Since all but one phase are fixed, the changes in the random trajectories
$\overrightarrow{x}(t,\theta)$ defined by the variation of $\theta
(\overrightarrow{k}_{o},\lambda_{o})$ in the integration should be vanishingly
small and then $\overrightarrow{x}(t,\theta)$ can be considered as a constant
in the integration. Thus the contribution of each mode to the expression of
$Q^{i}$ in (\ref{condit1}) vanishes.

In conclusion, the conditions for the validity of the translation invariance
of the stochastic forces are obeyed
\begin{equation}
{\Large \langle}Q^{i}(\overrightarrow{x}(t,\theta),\overrightarrow{v}%
(t,\theta)){\LARGE \rangle}_{\theta}=0{\Large ,}\nonumber
\end{equation}
and these relations imply the following drift momentum conservation laws
\begin{equation}
\frac{d}{dt}{\Large \langle}\mathcal{J}\text{ }^{i}\mathcal{(}%
\overrightarrow{x}(t,\theta),\overrightarrow{v}(t,\theta)){\LARGE \rangle
}_{\theta}=\frac{d}{dt}\ {\Large \langle}\frac{m\text{ }v^{i}(t,\theta)}%
{\sqrt{1-(\overrightarrow{v}(t,\theta))^{2}}}{\LARGE \rangle}_{\theta}=0.
\end{equation}

They indicate that the movement of a particle driven by these stochastic
forces can show a constant mean momentum, as it should be in the case that the
SQED theory is Lorentz invariant \cite{marshall,Boyer}.

\section{Covariant random forces}

\ In reference \cite{Alejandro} it was studied the action a covariantly
defined random force, by giving its expression in the rest system of a
particle as a function of the proper time. Numerically, it was shown that the
system can exhibit conserved drift velocities in spite of its random oscillation.
The problem was discussed in one time and one space dimension, but the
analysis might be extended to four dimensions. In this section, let us discuss
this problem in the\ 2D Minkowski metric in which the components of the vector
potential, the position vector and the metric are defined as
\begin{align}
x_{\mu}  &  =(x_{1},\text{ }x_{2})=(x_{1},i\text{ }x_{_{0}}),..\ g_{\mu\nu
}=\delta_{\mu\nu}\\
c  &  =1,\text{ \ \ \ \ }\ t=x_{_{0}},\nonumber
\end{align}
where the imaginary variable $x_{2}$, is analogous to the $x_{4}$ in four
dimensional space. Let us consider the equation of motion of the particle,
which in the laboratory frame is driven by a vacuum force defined by its value
in the rest system of the particle, as a function of the proper time. This
equation of motion can be written in the form
\begin{align}
\frac{dp^{\mu}(\tau)}{d\tau}  &  =\frac{1}{i}\epsilon^{\mu\nu}\frac{p^{\nu
}(\tau)}{m}\mathcal{F(\tau)},\label{newton}\\
\epsilon^{\mu\nu}  &  =\left\{
\begin{tabular}
[c]{l}%
$\ 0,$\ \ $\mu=\nu,$\\
$\ 1,\ $\ $\ \mu=1,$ $\nu=2,$\\
$-1,$ $\ \ \mu=2,$ $\nu=1,$%
\end{tabular}
\ \ \ \right.
\end{align}
in which $\epsilon^{\mu\nu}$ is the fully anti-symmetric tensor in two
dimensions, and the appearing 2D-momentum, 2D-velocity and the proper time of
the particle as a function of the usual time, are defined as
\begin{align}
p^{\mu}(\tau)  &  =m\text{ }u^{\mu}(\tau),\\
u^{\mu}(\tau)  &  =\frac{dx^{\mu}(\tau)}{d\tau},\\
\tau(t)  &  =\int_{-\infty}^{\infty}dt\sqrt{-\frac{dx^{\mu}(\tau)}{dt}%
\frac{dx^{\mu}(\tau)}{dt}}\nonumber\\
&  =\int_{-\infty}^{\infty}dt\sqrt{1-v_{1}^{2}(t)},\label{propert}\\
v_{1}(t)  &  =\frac{dx_{1}(t)}{dt}.
\end{align}

The stochastic nature of the problem is introduced by the random scalar
function of the proper time $\mathcal{F(\tau},\theta)\mathcal{). }$ It is
defined as a harmonic function of the proper time depending of the stochastic
parameter $\theta$ in a similar way as in the past section, in the form
\begin{equation}
\mathcal{F(\tau},\theta\mathcal{)=}\sum_{n=-\infty}^{\infty}\frac{2\pi}{T}%
\exp(-i(w_{n}\tau+\theta_{w_{n}}))S(w_{n}), \label{stoch1}%
\end{equation}
where $T$ is a large proper time interval to be taken arbitrary large but
finite, and the mode frequencies $\{w_{n}\}$ are defined as
\begin{equation}
w_{n}=\frac{2\pi}{T}n,\text{ \ }n=-\infty,...-1,0,1,...\infty. \label{stoch2}%
\end{equation}

Further, the function $S(w_{n})$ specifies a spectral weight for the proper
time frequencies in which the force is expanded. Finally, for each frequency
$\ w_{n}$ the stochastic element in the force is introduced in a similar form
as it is in SQED: that is, by adding aleatory phase functions $\theta_{w_{n}}$
in the argument of each harmonic function, which is uniformly distributed in
the interval $(-\pi,\pi).$ Again, the stochastic parameter $\theta$ is defined
as the set $\theta=\{\theta_{w_{n}}\}$ of all the stochastic phases (one for
each mode frequency $w_{n}$) and the probability density and the volume
differential in the space of parameters as
\begin{align}
\rho(\theta)  &  =\prod_{(w_{n})}\frac{\Theta(\pi-\theta_{w_{n}}%
))\Theta(\theta_{w_{n}}-\pi)}{2\pi},\nonumber\\
d\text{ }\theta &  =\prod_{(w_{n})}d\text{ }\theta_{w_{n}}. \label{stoch3}%
\end{align}

For the mean value of any function of the coordinates and velocities taken at
a fixed laboratory frame time $t$ we have%
\begin{align}
&  \langle\text{ }R(x_{1}(t,\theta)|\text{ }\overset{\cdot}{x_{1}}%
(t,\theta))\text{ }\rangle_{_{\theta}}\nonumber\\
&  =\int d\theta\text{ }\rho(\theta)\text{ }R(x_{1}(t,\theta)|\text{
}\overset{\cdot}{x_{1}}(t,\theta))\text{,}\nonumber\\
&  =\prod_{(w_{n})}\left(  \int_{-\pi}^{\pi}d\text{ }\theta(w_{n})\frac
{\Theta(\pi-\theta_{w_{n}}))\Theta(\theta_{w_{n}}-\pi)}{2\pi}\right)
\times\nonumber\\
&  R(x_{1}(t,\theta)|\text{ }\overset{\cdot}{x_{1}}(t,\theta)). \label{stoch4}%
\end{align}

\subsection{Lagrangian formulation}

Since the problem is relativistically invariant the Lagrangian formulation
becomes constrained, as it can be seen by contracting equation (\ref{newton})
with the 2D-momentum, which gives%

\begin{align}
p^{\mu}(\tau)\frac{dp^{\mu}(\tau)}{d\tau}  &  =\frac{1}{i}\epsilon^{\mu\nu
}\frac{p^{\mu}(\tau)p^{\nu}(\tau)}{m}\mathcal{F(\tau)}\\
&  =0=\frac{d\text{ }p^{2}(\tau)}{d\tau},\nonumber\\
p^{2}(\tau)  &  =ctc.
\end{align}
\bigskip

In order to find an unconstrained Lagrangian for the description of the
system, we can consider the single spatial component of the equations
(\ref{newton}), which can be written as
\begin{align}
\frac{dp^{1}(\tau)}{d\tau}  &  =\frac{1}{i}\epsilon^{12}\frac{p^{2}(\tau)}%
{m}\mathcal{F(\tau)},\\
\frac{dp^{1}(t)}{\sqrt{1-v^{2}}dt}  &  =\frac{1}{\sqrt{1-v^{2}}}%
\mathcal{F(\tau)},\\
\frac{d}{dt}\frac{m\text{ }v_{1}(t)}{\sqrt{1-v_{1}^{2}}}  &  =\mathcal{F(\tau
)},\\
v_{1}(t)  &  =\frac{dx_{1}(t)}{dt}.
\end{align}

The solution of this equation implies the satisfaction of the covariant ones
(\ref{newton}). Thus the full dynamics is defined by it. But, this single
equation can be derived from the Lagrangian
\begin{align}
L_{total}(x(t,\theta),\frac{dx(t,\theta)}{dt})  &  =-m\sqrt{1-(\frac
{dx(t,\theta)}{dt})^{2}}-\frac{dx(t,\theta)}{dt}\int_{-\infty}^{t}dt^{\prime
}\mathcal{F(\tau(}t^{\prime})\mathcal{)}\\
&  =L(x(t,\theta),\frac{dx(t,\theta)}{dt})+L_{ram}(x(t,\theta),\frac
{dx(t,\theta)}{dt}),
\end{align}
where $\mathcal{\tau(}t)$ is the proper time as a function of the laboratory
frame time $t$ given in (\ref{propert})$.$ Note in the above expression and in
what follows in this section, that we have redefined the spatial coordinate of
the particle by omitting the subindex $1$ as
\begin{equation}
x(t,\theta)\equiv x_{1}(t,\theta),
\end{equation}
in order to simplify the notation. Then, after considering the \ randomly
generated trajectories $\{x(t,\theta)\}$ solving the Lagrange equation of
motion%
\begin{equation}
\frac{d}{dt}\frac{m\text{ }\frac{dx(t,\theta)}{dt}}{\sqrt{1-(\frac
{dx(t,\theta)}{dt})^{2}}}=\mathcal{F(\tau(}t,\theta\mathcal{)),}%
\end{equation}
and considering that the Lagrangian is independent of the space translations
$x\rightarrow x+\delta a$, the stochastically conserved Noether current
(\ref{lagrancurr}), which can exist in the case that the stochastic forces also obey their
proper invariance condition (\ref{conditions}), has the form
\begin{align}
\mathcal{J}\text{ }^{r}\mathcal{(}x(t,\theta),\overset{\cdot}{x}(t,\theta))
&  =\frac{\partial L\mathcal{(}x(t,\theta),\overset{\cdot}{x}(t,\theta
)\mathcal{)}}{\partial\overset{\cdot}{x}(t,\theta)}\nonumber\\
&  =\frac{m\text{ }\frac{dx(t,\theta)}{dt}}{\sqrt{1-(\frac{dx(t,\theta)}%
{dt})^{2}}}.
\end{align}

Therefore, let us inspect the satisfaction of the invariance condition for the
stochastic forces, which in this case takes the form
\begin{align}
&  {\LARGE \langle}(\frac{\partial L^{ram}}{\partial x(t,\theta)}-\frac{d}%
{dt}\frac{\partial L^{ram}}{\partial\overset{\cdot}{x}(t,\theta)}%
){\LARGE \rangle_{_{\theta}}}\nonumber\\
&  ={\LARGE \langle}\mathcal{F}(\tau(t,\theta)){\LARGE \rangle}_{\theta}=0.
\end{align}

Now, from the definition of the stochastic force, its mean value can be
written in the form
\begin{align}
{\Large \langle}\mathcal{F(\tau(}t,\theta)\mathcal{)}{\LARGE \rangle}%
_{\theta}  &  =\prod_{(w_{n})}\left(  \int_{-\pi}^{\pi}d\text{ }\theta
(w_{n})\frac{\Theta(\pi-\theta_{w_{n}}))\Theta(\theta_{w_{n}}-\pi)}{2\pi
}\right)  \times\nonumber\\
&  \sum_{n^{\prime}=-\infty}^{\infty}\frac{2\pi}{T}\exp(-i(w_{n^{\prime}}%
\tau\mathcal{(}t,\theta)+\theta_{w_{n^{\prime}}}))G(w_{n^{\prime}}).
\end{align}

Then, similarly as it was done in the previous section, it is then possible to
consider a differential region in the stochastic parameter space, taken around
a fixed well defined values of the large number of the phase parameters
associated to each mode $(\theta_{w_{1}}....\theta_{w_{n}}...).$ and integrate
it over the values of a single mode phase factor $\theta_{w_{n^{\prime}}}$
over its region of definition $(-\pi,\pi).$ The contribution of a single mode
$w_{n^{\prime}}$ to this quantity leads to the expression
\begin{align}
&  =\prod_{(w_{n}\neq w_{n^{\prime}})}\left(  d\text{ }\theta(w_{n}%
)\frac{\Theta(\pi-\theta_{w_{n}}))\Theta(\theta_{w_{n}}-\pi)}{2\pi}\right)
\times\nonumber\\
&  \frac{2\pi}{T}\int_{-\pi}^{\pi}\theta_{w_{n^{^{\prime}}}}\exp
(-i(w_{n^{^{\prime}}}\tau\mathcal{(}t,\theta)+\theta_{w_{n^{\prime}}%
}))G(w_{n^{\prime}}).
\end{align}

Note that the whole mean values is a summation over $n^{\prime}$ and the
integration over the rest of the large number of phase factors of such terms.
Now, it can be noted that varying the phase of a single mode (within the large
number of them approximating a continuum) should change in a differential form
the expression of the proper time as a function of the laboratory frame time
$\tau\mathcal{(}t,\theta).$ Therefore, this function can be assumed as being
constant in the integration over $\theta_{w_{n^{\prime}}}$. This property in
turns implies the vanishing the mean value%
\begin{equation}
{\Large \langle}\mathcal{F(\tau(}t,\theta)\mathcal{)}{\LARGE \rangle_{\theta}%
}=0.
\end{equation}

Therefore, the satisfaction of the condition for the stochastic invariance of
the random forces implies the conservation of the linear momentum%
\begin{equation}
\frac{d}{dt}\ {\Large \langle}\frac{m\text{ }\frac{dx(t,\theta)}{dt}}%
{\sqrt{1-(\frac{dx(t,\theta)}{dt})^{2}}}{\Large \rangle}_{\theta}=0,
\end{equation}
a conservation law, which due to the Lorentz covariance of the discussion also
implies the energy constancy with time
\begin{equation}
\frac{d}{dt}\ {\Large \langle}\frac{1}{\sqrt{1-(\frac{dx(t,\theta)}{dt})^{2}}%
}{\Large \rangle_{\theta}}=0.
\end{equation}

Thus, the time invariance of a drift four-momentum is implied by the
stochastic form the Noether theorem also in this case.

\section*{Summary}

A generalized version of the Noether theorem is proven which extends  this
property for Lagrangian systems in which stochastic forces are present. It
shows that there are conserved in time functions of the coordinates and
velocities, when the stochastic forces have symmetries which coincide with the
invariances of the unperturbed Lagrangian (the non stochastic part of the
Lagrangian). The considered forces are of general kind including coordinate as
well as velocity dependencies. A characterization of the meaning of the
symmetry of the random forces and the definition of the stochastic mean values
being appropriate to formulate  the conservation laws, are also given. The
symmetry of the random forces are defined by the invariance of the moment
generating functionals of their random Lagrangian with respect to a symmetry
change of variables. The expression for the conserved Noether charges are
derived. The results are applied to the equation of motion of the particles in
Stochastic Quantum Electrodynamics \cite{Boyer,DelaP} and to the covariant
random forces discussed in \cite{Alejandro}. In both cases it follows that the
mean value over the ensemble of the relativistic spatial momentum of the
particles conserves in time \cite{DelaP}.

The analysis also indicates the possibility that two random classical systems
which interact during a finite time lapse, and afterwards flight far a apart,
can retain correlations between their physical quantities. This property may
describe apparent non causal correlations, existing between physical
quantities measured in the two well separated classical systems after they
stop to interact. This property, opens possibilities for the description of
EPR effects (\cite{EPR}) in hidden variable theories, being able to satisfy
the Bell's constraints about the existence of such theories \cite{bell}.

\ The presentation also suggests the interest in coupling a "self-field" to
stochastically driven particles as the ones discussed here. Assuming that a
momentum conservation might be derived%
%TCIMACRO{\U{b4} }%
%BeginExpansion
\'{}
%EndExpansion
for these systems, and also given that a reference frame exists in which the
momentum vanishes, the random distribution could describe spatially localized
stationary movements of the particle coupled with the self-field wave modes.
Then, if the system is relativistic invariant, in uniformly moving Lorentz
frames localized random structures will exists showing constant drift
velocities. This situation is quite resembling the one in the mentioned
Couder's experiments \cite{Couder1,Couder2,Couder3}. Therefore, these expected
particle wave composite structures have the chance of showing interference
effects when approaching to two slits in a wall. Such a possibility suggests a
connection of the Couder's experimental mechanical results with microscopic
hidden variable theories. We expect to consider these possibilities in
extensions of this work.

\section*{Acknowledgments}

The support granted by the N-35 OEA Network of the ICTP is greatly appreciated.


\begin{thebibliography}{99}                                                                                               %


\bibitem {Bogo}N. N. Bogoliubov, D. V. Shirkov, Introduction to the theory of
quantized fields, Editorial Nauka (1980).

\bibitem {Goldstein}H. Goldstein, C. Poole and J. Safko, Classical mechanics,
Addison-Wesley, Third Edition (2000).

\bibitem {Sokolov}A. A. Sokolov, I. M. Ternov, Electrodinamica euantica, Ed.
Mir Moscu (1989).

\bibitem {Landau}L. D. Landau, E. M. Lifshitz, Teoria clasica de campos, Vol
2, Ch. 10. (1992).

\bibitem {marshall}T. W. Marshall, Proc. R. Soc. A276, 475 (1963).

\bibitem {Boyer}T. Boyer, Phys.Rev. D 11, 790 (1975).

\bibitem {DelaP}L. de La Pe\~{n}a, A. M. Cetto, The quantum dice, ISBN
978-94-015-8723-5 (eBook), Alwyn van der Merwe, (1996).

\bibitem {Couder1}Y. Couder, S. Protiere, \ E. Fort and A. Boudaoud, \ Nature
437 ,208 (2005).

\bibitem {Couder2}S. Protiere, A. Boudaoud and Y. Couder, Fluid Mech. 554, 85 (2006).

\bibitem {Couder3}A. Eddi, E. Fort, F. Moisy and Y. Couder, Phys. Rev. Lett.
102, 240401 (2009).

\bibitem {baez}J. Baez and B. Fong, \ J. Math. Phys. 54 , 013301 \ (2013). arXiv:1203.2035.

\bibitem {Alejandro}A. Cabo-Bizet, A. Cabo Montes de Oca, Phys. Lett. A. 359,
265 (2006).

\bibitem {Vankampen}N. G. van Kampen, Stochastics processes in physics and
chemistry, Elsevier, Third Edition (2007).

\bibitem {Caceres}M. O. C{\'{a}}ceres, Elementos de estad{\'{\i}}stica del no
equilibrio. y procesos estoc{\'{a}}sticos, Editorial Revert\`{e}\ (2003).

\bibitem {bell}J. S. Bell, Rev. Mod. Phys. 38, 447 (1966).

\bibitem {EPR}A. Einstein, B. Podolsky and N. Rosen, Phys. Rev. 47, 777 (1935).
\end{thebibliography}
\end{document}